\documentclass[a4paper]{jpconf}
\usepackage{graphicx}

\def\urs{URu$_2$Si$_2$}

\usepackage{bm}
\usepackage{epsfig}

\begin{document}
\title{Inelastic contribution of the resistivity in the hidden order in \urs}

\author{E Hassinger, T D Matsuda\footnote[1]{Present address:
Advanced Science Research Center, Japan Atomic Energy Agency, Ibaraki 319-1195, Japan}, G Knebel, V Taufour, D Aoki and J Flouquet}
\address{INAC-SPSMS, CEA Grenoble, 17 rue des Martyrs, 38054 Grenoble, France}

\ead{georg.knebel@cea.fr}

\begin{abstract}
In the hidden order of \urs\ the resistivity at very low temperature shows no $T^2$ behavior above the transition to superconductivity. However, when entering the antiferromagnetic phase, the Fermi liquid behavior is recovered. We discuss the change of the inelastic term when entering the AF phase with pressure considering the temperature dependence of the Gr\"{u}neisen parameter at ambient pressure and the influence of superconductivity by an extrapolation of high field data.
\end{abstract}

\section{Introduction}
\urs\ has attracted the interest of researchers for a long time, because it has an ordered phase below $T_0 = 17.8$\,K with huge anomalies in macroscopic quantities at the transition but unknown order parameter, which coexists with unconventional superconductivity below $T_{SC}= 1.5$\,K \cite{Palstra1985, Maple1986}. Its resistivity is anisotropic and by a factor of  about two larger in the basal plane of the tetragonal crystal structure than along the crystallographic $c$ axis. At the transition to the so-called hidden order (HO) at $T_0$, the resistivity shows a nesting like behavior with a jump due to partial gapping of the Fermi surface and a loss of carriers by a factor of 3-10 \cite{Maple1986,Dawson1989} which has also been determined by band structure calculations \cite{Harima2010}. By angular resolved photoemission spectroscopy it was directly shown that when entering the HO, some bands drop below the Fermi energy \cite{Santander2009}. There are signs of different kinds of gaps (spin gap and charge gap) in many properties observable by an exponential behavior of macroscopic quantities \cite{Palstra1985, Maple1986}, gap in optical conductivity \cite{Bonn1988} and sharply gapped Ising-like longitudinal magnetic excitations in neutron scattering \cite{Broholm1987, Broholm1991}. The small ordered magnetic moment below $T_0$ was proven to be extrinsic \cite{Matsuda2003}. Recent theories tend to describe the hidden order by a local multipole order parameter \cite{Haule2009,Harima2010b}, but have to take into account the change of hybridization of the conduction electrons \cite{Schmidt2010}. Certainly, both effects are mixed and play a role in the formation of the HO phase. Under pressure, the ground state changes from HO to an antiferromagnetic phase (AF) at $P_x = 0.5$\,GPa and superconductivity is suppressed at the same time \cite{Amitsuka1999,Amitsuka2007,Hassinger2008}. However, the critical pressure $P_x$ seems to depend strongly on the sample and pressure homogeneity conditions \cite{Motoyama2003,Amitsuka2008}. The resistivity anomaly does not change at the transition to either phase below $T_0$ or $T_N$, indicating that the rearrangement of the Fermi surface is similar in both phases. 
Below $T_0$ or $T_N$, the resistivity can be described quite well by the following formular \cite{Palstra1986}
\[\rho (T) = \rho_0 + AT^2 + B\frac{T}{\Delta}(1+2\frac{2T}{\Delta})\exp(-\frac{\Delta}{T})\]
which describes the resistivity of a Fermi liquid interacting with gapped spin waves in an antiferromagnet \cite{Hessel-Andersen1980}. Obviously, this is not the right model in the non-magnetic HO state as described in Ref. \cite{Hassinger2008}. Nevertheless, it allows to determine the value of the anisotropic gap $\Delta$, which increases by 20\,\% when entering the AF phase \cite{Aoki2009}.
Already McElfresh {\it et al.} \cite{McElfresh1987} have remarked, that at low temperature one never finds a $T^2$ behavior, even for a very small temperature range just above the superconducting transition. Recently, an anisotropic inelastic term of the resistivity in the HO state was found \cite{Zhu2009} which could indicate an anisotropic scattering mechanism due to the order parameter in this phase. In Ref.\cite{Miyake2010} it has been shown that a quadrupolar charge order and its excitation can cause this anisotropy. However, a detailed study on crystals with different quality should clarify the influence of sample quality \cite{Matsuda2010}.
If this anisotropy is a signature of the special order parameter of the HO state, the temperature behavior of the resistivity should change under pressure when entering the antiferromagnetic state.
Here we present a high pressure resistivity experiments to address this question. 

\section{Experimental Details}

The single crystal was grown by Czochralsky pulling method in a tetra arc furnace. We have carried out resistivity measurements with current along the $a$-axis and magnetic field along $c$axis  under pressure.  Due to the non-quadratic temperature dependence of the resistivity at low temperature just above the superconducting transition \cite{McElfresh1987,Hassinger2008}, the extrapolation to $T = 0$ is ambiguous and can lead to a nearly zero or even negative residual resistivity in very good crystals. Therefore, we defined the residual resistivity ratio RRR as $\rho(300K)/\rho(2K)$ above the onset of the superconducting transition. In this study, the crystal had RRR = 160. We have carried out standard four point resistivity measurements with a current of $I \approx 10$\,$\mu$A through the sample. The signal was amplified by a low temperature transformer by factor of 1000 keeping the noise level very low. The measurement frequency of 71\,Hz was chosen because of its bandpass. 
The signal was further amplified by a room temperature pre-amplifier and detected with a lock-in amplifier. Low temperatures were created with a dilution refrigerator with base temperature of about 20\,mK. 
Pressure was applied with a copper beryllium piston cylinder pressure cell. 
For the determination of the upper critical field $H_{c2}$, field sweeps at constant temperature were carried out at low temperatures. $H_{c2}$ is defined where the resistivity starts to rise from zero.

\begin{figure}
\begin{minipage}[b]{74mm}
\begin{center}
\includegraphics[width=70mm]{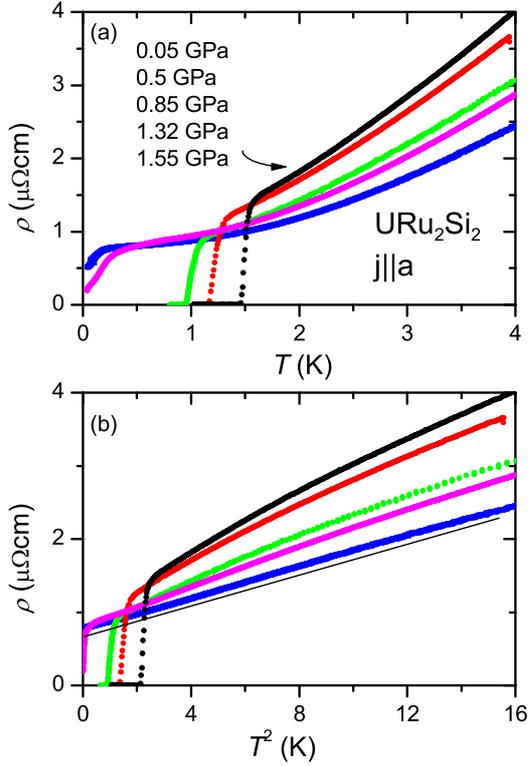}
\end{center}
\end{minipage}
\hspace{5mm}
\begin{minipage}[b]{74mm}
\caption{\label{Tdeprho}(Color online) (a) Resistivity as a function of temperature in zero magnetic field for several pressures.  (b) The same data as a function of $T^2$. The straight line is a guide to the eye.}
\end{minipage}
\end{figure}

\section{Results and Discussion}

The temperature dependence of the resistivity for the measured pressures is shown in Fig.~\ref{Tdeprho} as a function of $T$ (top) and $T^2$ (bottom). One clearly sees, that the resistivity does not show a $T^2$ dependence over an extended temperature range for the low pressures whereas for $P=1.55$\,GPa a nice $T^2$ up to 3.5\,K is observed.  Only for the three lowest pressures a complete superconducting transition with zero resistivity appears whereas only an onset of the superconducting transition if observed for the highest pressures.

\begin{figure}
\begin{minipage}[b]{74mm}
\begin{center}
\includegraphics[width=70mm]{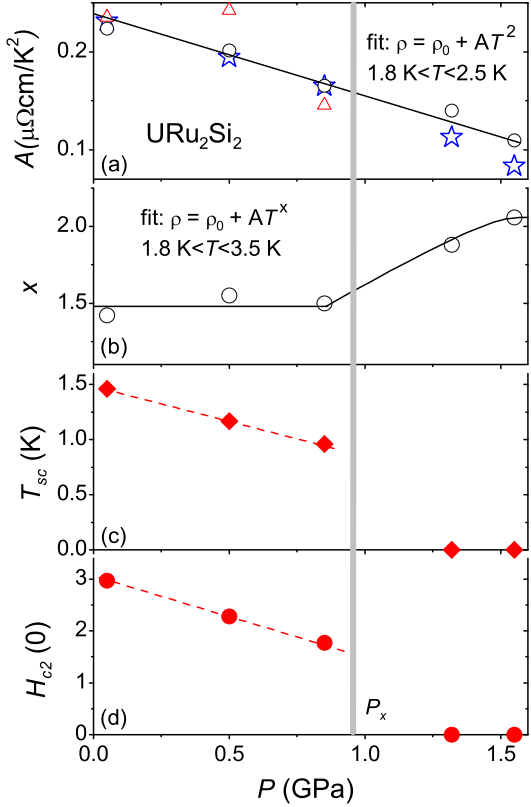}
\end{center}
\end{minipage}
\hspace{5mm}
\begin{minipage}[b]{74mm}
\caption{\label{param}Pressure dependence of several parameters determined from resistivity measurements at zero field (a-c) and with $H \parallel c$. In (a) we further show the pressure dependence of the normalized square of the effective mass of the heavy $\beta$ branch (blue stars) and the normalized square of the initial slpope of the upper critical field (triangles). The critical pressure $P_x$ is indicated by the gray line. Lines are to guide the eye, indicating a $T^2$ dependence. }
\end{minipage}
\end{figure}

In Fig.~\ref{param} we plotted the pressure dependence of some parameters one can extract from these data. One striking feature in almost all of them is the discontinuous pressure dependence between 0.85\,GPa and 1.32\,GPa. It evidently shows, that the transition between the HO and the pressure induced AF state appears between these two pressures, i.e. $0.85$\,GPa\,$<P_x<1.32$\,GPa. Magnetoresistance measurements on the same sample also show a clear change between the lower three and the higher two pressures \cite{Hassinger2010}. In Fig.~\ref{param}a, we show the $A$ coefficient of a forced $\rho = \rho_0 + AT^2$ fit at $H = 0$ between 1.8\,K and 2.5\,K. It decreases strongly with pressure. However, in the present measurements, the pressure steps were too large to detect an discontinuity of $A$ at the critical pressure $P_x$ \cite{Hassinger2008}.
As $A\propto (m^{\star})^2$ we have plotted the normalized square of the effective band mass of the $\beta$ branch determined in recent Shubnikov-de Haas measurements \cite{Hassinger2010} (blue stars) as well as the normalized square of the initial slope of the upper critical field $H_{c2}$ near $T_{SC}$ (small triangles), which are both proportional to the effective mass as a function of pressure. The pressure dependence of all these parameters is consistent. 

As expected from a bare look at the data, this $T^2$ fit does not reproduce the data well at low pressures. Therefore, we fitted the data also with a $\rho = \rho_0 + AT^x$ fit for temperatures between 1.8\,K and 3.5\,K (Fig.~\ref{param}b). The exponent is around 1.5 in the HO state and a clear $T^2$ behavior is found only deep inside the AF phase.
The superconducting parameters $T_{SC}$ and $H_{c2}$ (Fig.~\ref{param}c and d) decrease linearly with pressure and then disappear abruptly above $P_x$. The critical pressure where superconductivity would disappear extrapolated from the low pressure data is at around 2\,GPa. In earlier pressure studies of the resistivity on samples with lower quality, the decrease $T_c (P)$ was linear \cite{Hassinger2008,Jeffries2008, McElfresh1987} up to the pressure where the superconducting phase disappeared. The superconducting transition was found up to quite high pressures (far above 1\,GPa for the above mentioned references). In contrast, in magnetic susceptibility and especially specific heat measurements, which are sensitive to bulk transitions, the superconducting state disappeared at the critical pressure \cite{Amitsuka2007,Hassinger2008}. The explanation for this difference is that in the same way as droplets of the AF phase exist within the HO state ($P<P_x$) near lattice imperfections (and are responsible for the small moment observed in neutron scattering) droplets of the superconducting HO state exist under pressure ($P>P_x$) in the non-superconducting AF phase. In resistivity, a shortcut through the sample by inhomogeneous superconducting droplets can give the zero resistivity signal.

The non-Fermi liquid behavior in resistivity is observed precisely in the HO phase and disappears in the AF phase. The same happens with superconductivity. It is tempting to claim, that the component of linear temperature behavior is a signature of the HO state or even stronger, that superconductivity is linked to a linear resistivity term \cite{Doiron-Leyraud2009} as was claimed in high temperature and organic superconductors. However, in \urs\, one should be careful with claims like this.
A recent study of the Gr\"{u}neisen parameter gives a saturation only at very low temperature \cite{Hardy2010}. Accordingly, the Fermi liquid behavior in resistivity is maybe hidden by the onset of superconductivity. Zhu {\it et al.} have studied the temperature behavior of the resistivity in field with a longitudinal configuration ($H\parallel I$) to avoid transverse magnetoresistance. If the resistivity is expressed as $\rho_0 +AT^x$, they report $x<2$ for the zero field measurements in both crystal directions. With field, a Fermi liquid behavior is recovered above $H_{c2} = 3$\,T for a current and field along the $c$ axis whereas a linear temperature dependence for current and field along $a$ remains up to the maximum field of $H = 12$\,T. 
However this field is not high enough to suppress superconductivity along the $a$ axis completely. To overcome this problem we have performed magnetoresistance measurements as function of field for different constant temperatures \cite{Matsuda2010}. 
Extrapolating the magnetoresistance from the normal state at high fields above $H_{c2}$ to zero field 
at constant temperatures $T<T_{SC}$, the resistivity in the normal state in absence of superconductivity has been extrapolated. 
From this extrapolations a $T^2$ behavior \cite{Matsuda2010} with a coefficient $A$ in good agreement with our result at zero pressure has been observed. 

Knowing the Gr\"{u}neisen parameter $\Gamma=40$ \cite{Hardy2010} and the compressibility $\kappa\approx 0.5 \cdot 10^{-6}$\,bar$^{-1}$ \cite{Jeffries2010} allows a prediction for the relative change of $A$ with pressure:
As $A \propto (m^{\star})^2$ with the characteristic temperature $T^{\star} \propto 1/m^{\star}$, then $dT^{\star}/T^{\star}=dm^{\star}/m^{\star}=1/2dA/A$. 
With the definition of the Gr\"{u}neisen parameter \[ \Gamma = -d\ln{T^{\star}}/d\ln{V}=\kappa^{-1}d\ln{T^{\star}}/dp\]
we find for a pressure difference of $\Delta p = 0.5$\,GPa a relative change of $\Delta A/A \approx 0.2$ in agreement with our data.

\section{Conclusion}
To summarize, we show the change of the temperature dependence of the electrical resistivity in \urs\ with pressure. As expected from the Gr\"{u}neisen parameter, Fermi liquid behavior appears at very low temperatures $T<T_{SC}$ only. Superconductivity has to be suppressed to be able to observe a $T^2$ temperature dependence. However, we cannot exclude peculiar scattering in the HO state, but it is difficult to find a decisive answer.

\ack
We acknowledge funding from the French ANR within the projects CORMAT, DELICE, and SINUS. 

\section*{References}

\bibliography{urs}

\end{document}